\documentclass[twocolumn, iop, apj, twocolappendix, numberedappendix, appendixfloats]{emulateapj}
\usepackage{CJKutf8}
\usepackage{threeparttable}
\usepackage{multirow}
\usepackage{times}
\usepackage{enumitem}
\usepackage{url} 
\usepackage{color}
\usepackage{amsmath,bm}
\usepackage{subfigure}
\usepackage{natbib}
\usepackage{graphicx}
\usepackage{graphics}
\usepackage{hyperref}                             
\usepackage{amsmath}
\usepackage{bm}
\usepackage{subfigure}
\usepackage{array}
\usepackage{booktabs}
\usepackage{lineno}

\shorttitle{The warp and waves in the Milky Way disk}
\shortauthors{Sun et al.}
\submitted{Submitted to ApJ;\,\,\,\,Accepted April 24, 2025}

\begin{document}


\title{The Kinematic Signature of the Warp and Waves in the Milky Way Disk}

\author{Weixiang Sun\textsuperscript{1,5}}
\author{Han Shen\textsuperscript{2,3}}
\author{Biwei Jiang\textsuperscript{1,5}}
\author{Xiaowei Liu\textsuperscript{4,5}}

\altaffiltext{1}{School of Physics and Astronomy, Beijing Normal University, Beijing 100875, People’s Republic of China; {\it sunweixiang@bnu.edu.cn {\rm (WXS)}}; {\it bjiang@bnu.edu.cn {\rm (BWJ)}}}
\altaffiltext{2}{School of Physics, University of New South Wales, Kensington 2052, Australia}
\altaffiltext{3}{ARC Centre of Excellence for All Sky Astrophysics in 3 Dimensions (ASTRO 3D), Australia}
\altaffiltext{4}{South-Western Institute for Astronomy Research, Yunnan University, Kunming 650500, People's Republic of China; {\it x.liu@ynu.edu.cn {\rm (XWL)}}}
\altaffiltext{5}{Corresponding authors}

\begin{abstract}

Using over 170,000 red clump (RC) stars selected from LAMOST and APOGEE, we conduct a detailed analysis of the stellar $V_{Z}$ as a function of $L_{Z}$ (or $R_{g}$) across different $\phi$ bins for various disk populations.
The $V_{Z}$ of the whole RC sample stars exhibits a wave-like pattern superimposed on an exponentially increasing trend, indicating the contribution from disk warp, disk flare and disk waves.
Our results across various populations suggest that the thin disk is similar to the whole RC sample behavior, while the thick disk displays a wave-like pattern superimposed on a linearly increasing trend, meaning that the features of disk warp and waves are present in both thin and thick disks, and the disk flare feature is only present in the thin disk.
These results indicate that the disk warp is potentially driven by secular processes like disk perturbations from intergalactic magnetic fields and a misaligned dark halo.
The line-of-node of the disk warp of various populations displays a slight difference, with $\phi_{0}$ = 5.68 $\pm$ 2.91 degree for the whole RC sample stars, $\phi_{0}$ = 5.78 $\pm$ 2.89 degree for the thin disk stars, and $\phi_{0}$ = 4.10 $\pm$ 3.43 degree for the thick disk stars.

\end{abstract}


\section{Introduction}

The fact that the majority of spiral galaxies are strongly warped in their outer disk, is in good agreement with the results determined by vertical bending waves \citep[e.g.,][]{Hunter1969}.
As a typical spiral galaxy, the Milky Way has a well-known warped disk that has been widely confirmed by neutral gas \citep[e.g.,][]{Kerr1957}, molecular clouds disk \citep[e.g.,][]{Grabelsky1987}, interstellar dust \citep[e.g.,][]{Freudenreich1994, Drimmel2001}, as well as stars \citep[e.g.,][]{Chen2019, Poggio2018, Poggio2024, Sun2025}.
The results indicate that the Galactic disk is flat out to the Solar radius, then bends downwards in the south and upwards in the north, with the line-of-node (LON) close to the Galactic Center–Sun direction \citep[e.g.,][]{Lopez-Corredoira2002, Momany2006}.
Some studies also reveal that the outer disk is likely more complex than a simple warp, most of those point to a wave-like pattern \citep[e.g.,][]{Khanna2019, Friske2019, Antoja2022}, with amplitudes that can exceed $\sim$1.0\,kpc \citep[e.g.,][]{Xu2015, Price-Whelan2015}.

Several studies have been made to character the kinematic properties of the Galactic warp and waves based on the stellar vertical velocity ($V_{Z}$) distributions \citep[e.g.,][]{Gaia Collaboration2018, Huang2018, Cheng2020, Hunt2025}.
Most of those results indicate that the $V_{Z}$ of disk stars can be described as $V_{Z}$ = $b$ + $a\,L_{Z}$ + $A$\,sin(2$\pi$c/$L_{Z}$\,+\,$d$) \citep[e.g.,][]{Schonrich2018}, in which $b$ + $a\,L_{Z}$ reveals a simple, non-wrapped, perfectly static warp structure, where $L_{Z}$ is vertical angular momentum, defined as $L_{Z}$ = $RV_{\phi}$.
$A$\,sin(2$\pi$c/$L_{Z}$\,+\,$d$) displays a pattern of the disk that may oscillate vertically as radial propagating waves, where $A$, $c$ and $d$ are respectively the amplitude, the period and the phase of the wave-like pattern.

Sch{\"o}nrich \& Dehnen ({\color{blue}{2018}}) present a detailed analysis of the relation between $V_{Z}$ and $L_{Z}$ for the TGAS sample \citep[e.g.,][]{Lindegren2016}.
They find the $V_{Z}$ displays a global increasing trend with $L_{Z}$, consistent with expectations from a long-lived Galactic warp \citep[e.g.,][]{Drimmel2000}.
In addition, their results also revealed a wave-like pattern of $V_{Z}$ as a function of $L_{Z}$ that may be from a winding warp or bending waves.
These results were later independently supported by LAMOST-TGAS sample \citep[e.g.,][]{Huang2018}.

However, a detailed analysis of the relation of $V_{Z}$--$L_{Z}$ for different stellar populations on a larger disk volume is still not well characterized since the lack of accurate measurements of stellar kinematic parameters and stellar atmospheric parameters in the samples used in previous studies \citep[e.g.,][]{Schonrich2018, Huang2018}.
These observation limitations meant that the nature of the disk warp and waves are not yet well measured in a larger Galactic disk radius that would enable a credible assessment of their origins and whether they are long-lived features in the Milky Way disk.
At present, the larger sample of RC stars \citep[e.g.,][]{Bovy2014, Huang2020, Sun2024a} selected from LAMOST and APOGEE surveys, with combining stellar kinematic parameters from Gaia survey, presents an excellent opportunity to conduct research this field.
Leveraging this sample, we can conduct a comprehensive analysis of the $V_{Z}$--$L_{Z}$ relation across the Galactic disk, thereby gaining deeper insights into the disk structures and their evolutions.

This paper is structured as follows.
In Section\,2, we describe the data used in this work,
and present our results and discussion in Section\,3. 
Finally, our main conclusions are summarized in Section\,4.

\section{Data}

In this paper, we mainly used a sample with 171,320 RC stars from APOGEE \citep[e.g.,][]{Majewski2017} and the LAMOST \citep[e.g.,][]{Cui2012, Deng2012, Liu2014, Yuan2015} surveys, in which, 137,448 RC stars from the LAMOST and 39,675 RC stars from APOGEE.
The typical uncertainties of the surface gravity (log$_{g}$), effective temperature ($T_{\rm eff}$), line-of-sight velocity ($V_{\rm r}$), [Fe/H] and [$\alpha$/Fe], are respectively, 0.10\,dex, 100\,K, 5\,km s$^{-1}$, 0.10$-$0.15\,dex and 0.03$-$0.05\,dex \citep[e.g.,][]{Bovy2014, Sun2024a}.
Since the distance is determined from the standard-candle property of RC stars, their distance errors are generally smaller than 5\%--10\%.
To ensure the accurate kinematic calculation, we further update the stellar atmospheric parameters of the whole sample stars to Gaia DR3 \citep[e.g.,][]{Gaia Collaboration2023a, Gaia Collaboration2023b, Recio-Blanco2023}.

\begin{figure}[t]
\begin{center}
\includegraphics[width=8.3cm]{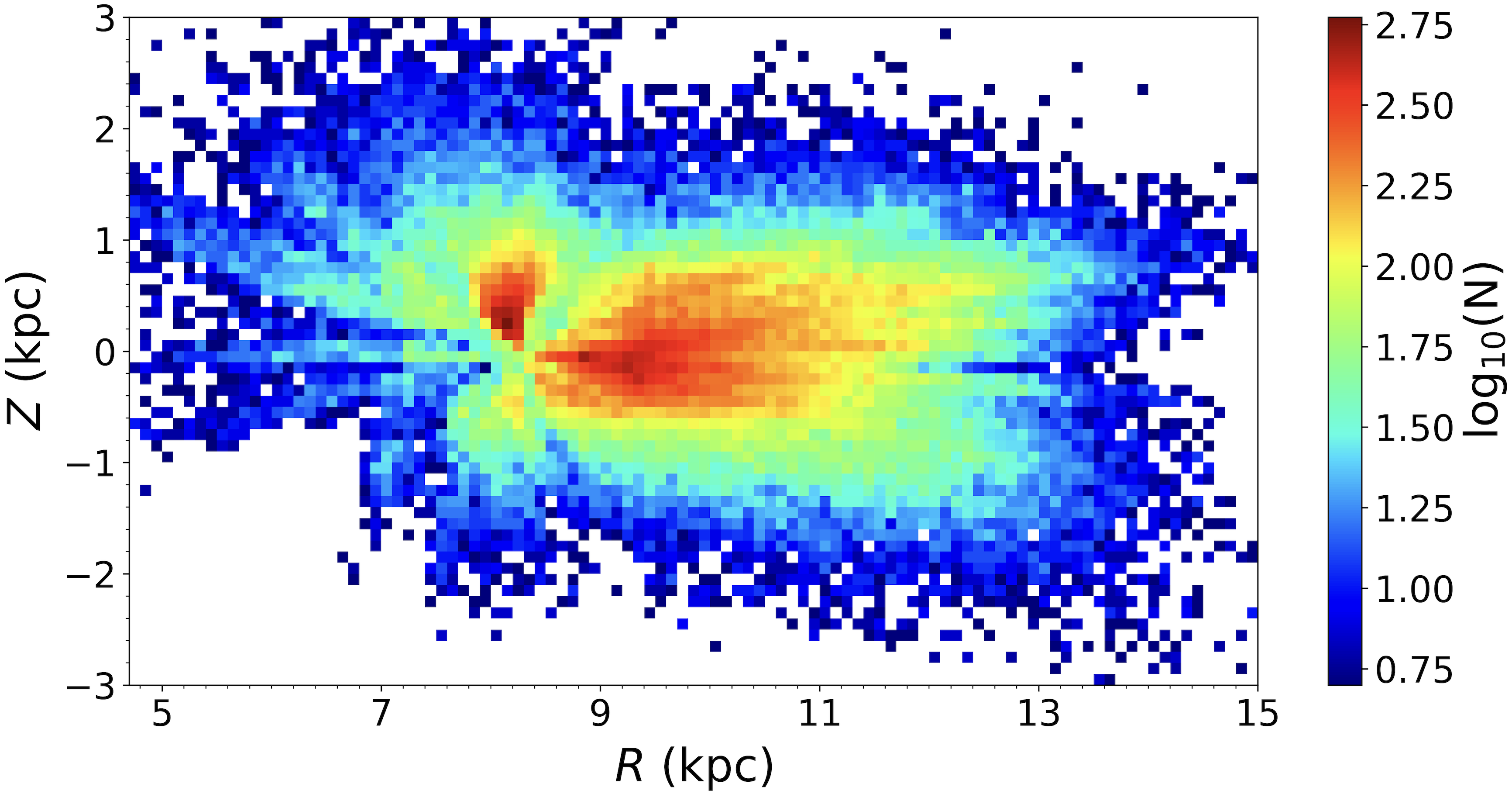}
\caption{Spatial distribution in the $R$ - $Z$ plane, of the sample stars, with color-coded by the stellar number density.
There is a minimum of 8 stars per bin spaced 0.1\,kpc in both axes.}
\end{center}
\end{figure}

In this paper, we used the standard Galactocentric cylindrical Coordinate ($R$, $\phi$, $Z$), with the $V_{R}$, $V_{\phi}$ and $V_{Z}$ are the three velocity components, respectively.
To estimate the 3D positions and 3D velocities, we adopt the Galactocentric distance of the Sun as $R_{\odot}$ = 8.34 kpc \citep{Reid2014}, the solar motions as ($U_{\odot}$, $V_{\odot}$, $W_{\odot}$) $=$ $(13.00, 12.24, 7.24)$ km s$^{-1}$ \citep{Schonrich2018}, and the local circular velocity as $V_{c,0}$ = 238 km s$^{-1}$ \citep[e.g.,][]{Reid2004, Schonrich2012, Bland-Hawthorn2016}.
Stellar orbital parameters, such as, vertical angular momentum $L_{Z}$ and guiding center radius $R_{g}$, are respectively, simple determined by $L_{Z}$ = $RV_{\phi}$ and $R_{g}$ = $RV_{\phi}$/$V_{c,0}$.

\begin{table*}

\caption{The properties of thin and thin disk populations.}

\centering
\setlength{\tabcolsep}{7mm}{
\resizebox{2.\columnwidth}{!}{
\begin{tabular}{lllllllll}
\hline
\hline
\specialrule{0em}{7pt}{0pt}
Name                                         &   $\left\langle R \right\rangle$   &   $\left\langle V_{\phi} \right\rangle$   &     $\sigma_{R}$        &    $\sigma_{\phi}$    &       $\sigma_{Z}$       &  $\left\langle V_{R}V_{Z} \right\rangle$  &     Number     &    n\,(ratio) \\[0.07cm]
                                             &                (kpc)               &               (km\,s$^{-1}$)              &     (km\,s$^{-1}$)      &     (km\,s$^{-1}$)    &      (km\,s$^{-1}$)      &          (km$^{2}$\,s$^{-2}$)             &                &                \\
\specialrule{0em}{7pt}{0pt}
\hline
\specialrule{0em}{7pt}{0pt}
Thin disk                                    &                 10.14              &                     225.01                &        33.86            &           22.97       &          19.19           &                17.98             &    135,009     &      79.08\%\\ [0.2cm]
Thick disk                                   &                 8.54               &                     179.27          &        60.61            &           54.81       &          37.03           &                120.84                     &    23,168      &      13.57\%   \\
\specialrule{0em}{7pt}{0pt}
\hline
\specialrule{0em}{7pt}{0pt}
\end{tabular}}
}
\label{tab:datasets}
\end{table*}

\begin{figure*}[t]
\centering

\subfigure{
\includegraphics[width=15.5cm]{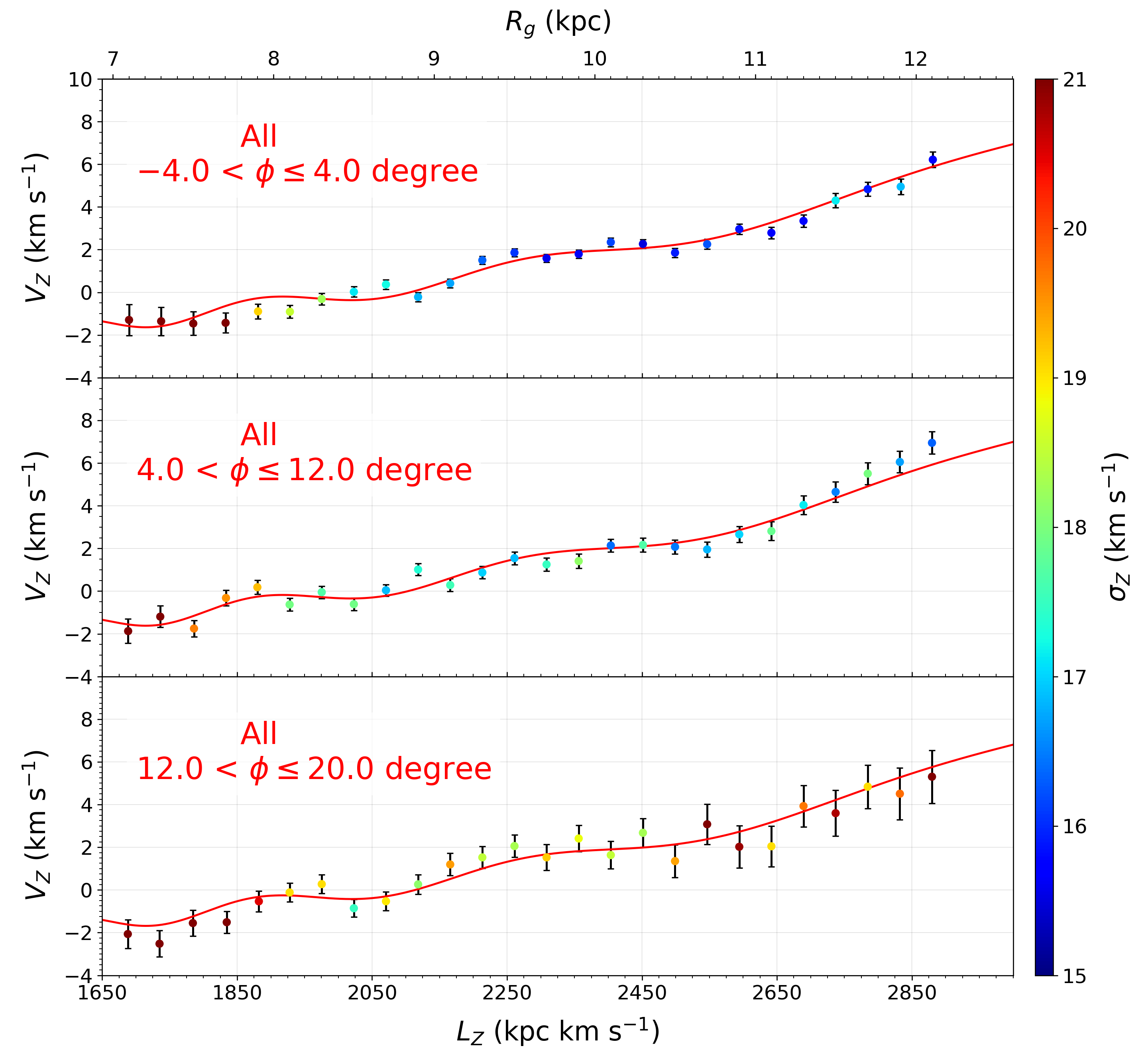}
}

\caption{The $V_{Z}$ as a function of $L_{Z}$ for the whole RC sample stars, color-coded by $\sigma_{Z}$, with no less than 30 stars in each bin.
The corresponding $R_{g}$ are also labeled at the top of the figure.
The red line is the best fit with Equation (3) for the relation of $V_{Z}$ -- $L_{Z}$.}
\end{figure*}

\begin{table*}
\caption{Parameters of Equation (3) are obtained for the whole RC sample stars and thin disk stars, and parameters of Equation (2) are obtained for the thick disk stars.}

\centering
\setlength{\tabcolsep}{4.3mm}{
\begin{tabular}{llllllll}
\hline
\hline
\specialrule{0em}{7pt}{0pt}
Name                                &     $b$        &      $a$                       &       $B$            & $\phi_{0}$  &      $A$       &       $c$            &  $d$       \\[0.07cm]

                                    & (km\,s$^{-1}$) & ($\times$10$^{-3}$ kpc$^{-1}$) & ($\times$10$^{-4}$)  &   (degree)  & (km\,s$^{-1}$) &  (kpc\,km\,s$^{-1}$) &             \\

\specialrule{0em}{7pt}{0pt}
\hline
\specialrule{0em}{7pt}{0pt} 
All stars                           &   $-$5.42\,$\pm$\,0.17 &           -                &  8.32\,$\pm$\,0.15 &   5.68\,$\pm$\,2.91   &   0.39\,$\pm$\,0.15   &   10705.62\,$\pm$\,425.35 &  $-$2.71\,$\pm$\,1.15     \\[0.3cm]
Thin disk                           &   $-$5.54\,$\pm$\,0.22 &           -                &  8.40\,$\pm$\,0.21 &   5.78\,$\pm$\,2.89   &   0.48\,$\pm$\,0.12   &   10543.33\,$\pm$\,354.88 &  $-$2.31\,$\pm$\,0.97     \\[0.3cm]
Thick disk                          &   $-$2.17\,$\pm$\,1.18 &    0.99\,$\pm$\,0.05       &        -           &   4.10\,$\pm$\,3.43   &   0.60\,$\pm$\,0.22   &   12563.03\,$\pm$\,639.98 &  $-$2.22\,$\pm$\,1.62     \\
\specialrule{0em}{7pt}{0pt}
\hline 
\specialrule{0em}{7pt}{0pt}
\end{tabular}}
\label{tab:datasets}
\end{table*}

\begin{figure*}[t]
\centering
\subfigure{
\includegraphics[width=15.5cm]{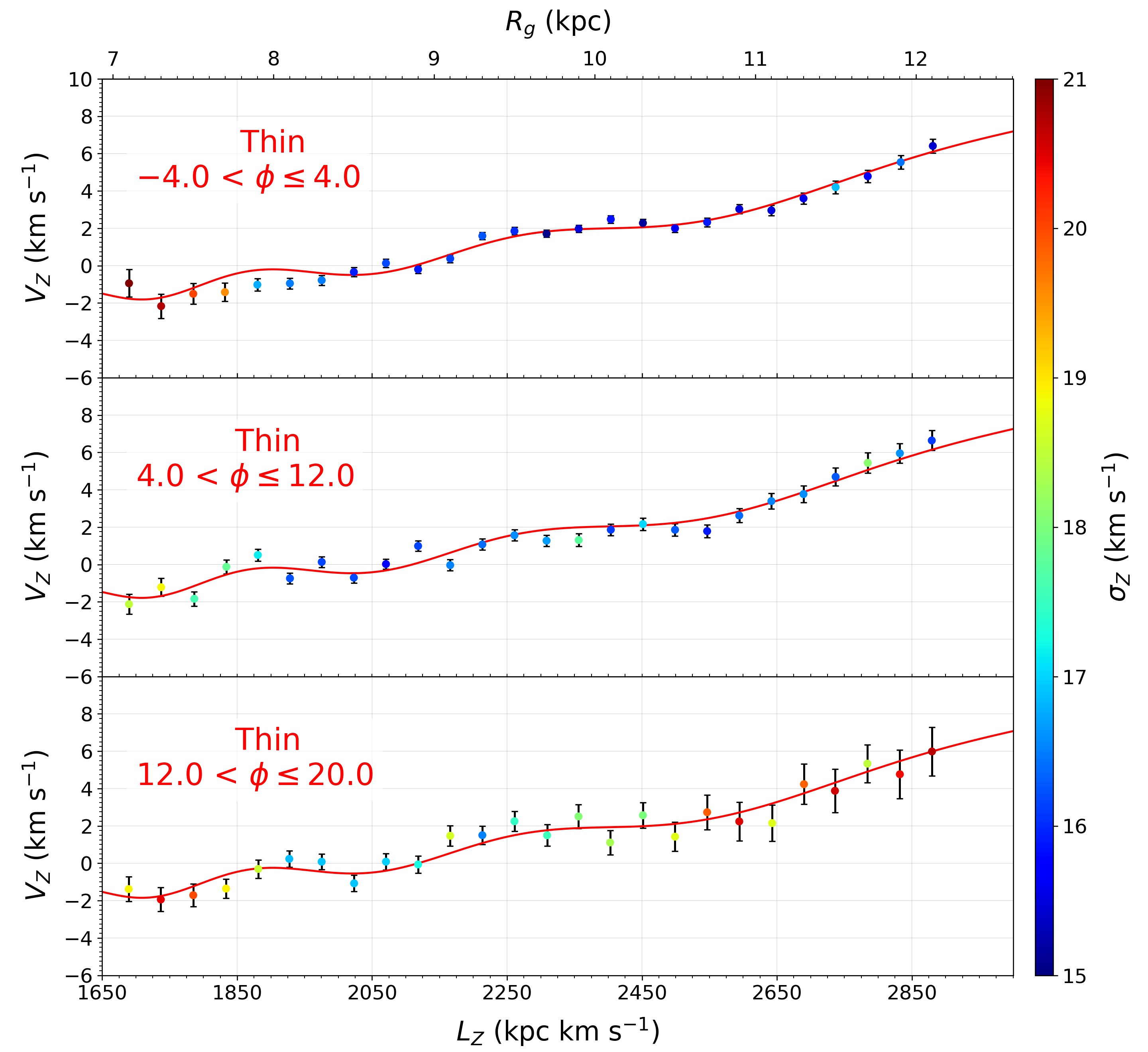}
}

\caption{Similar to Fig.\,2 but for the thin disk stars. The red lines are the best fits determined by Equation (3).}
\end{figure*}

\begin{figure*}[t]
\centering
\subfigure{
\includegraphics[width=15.5cm]{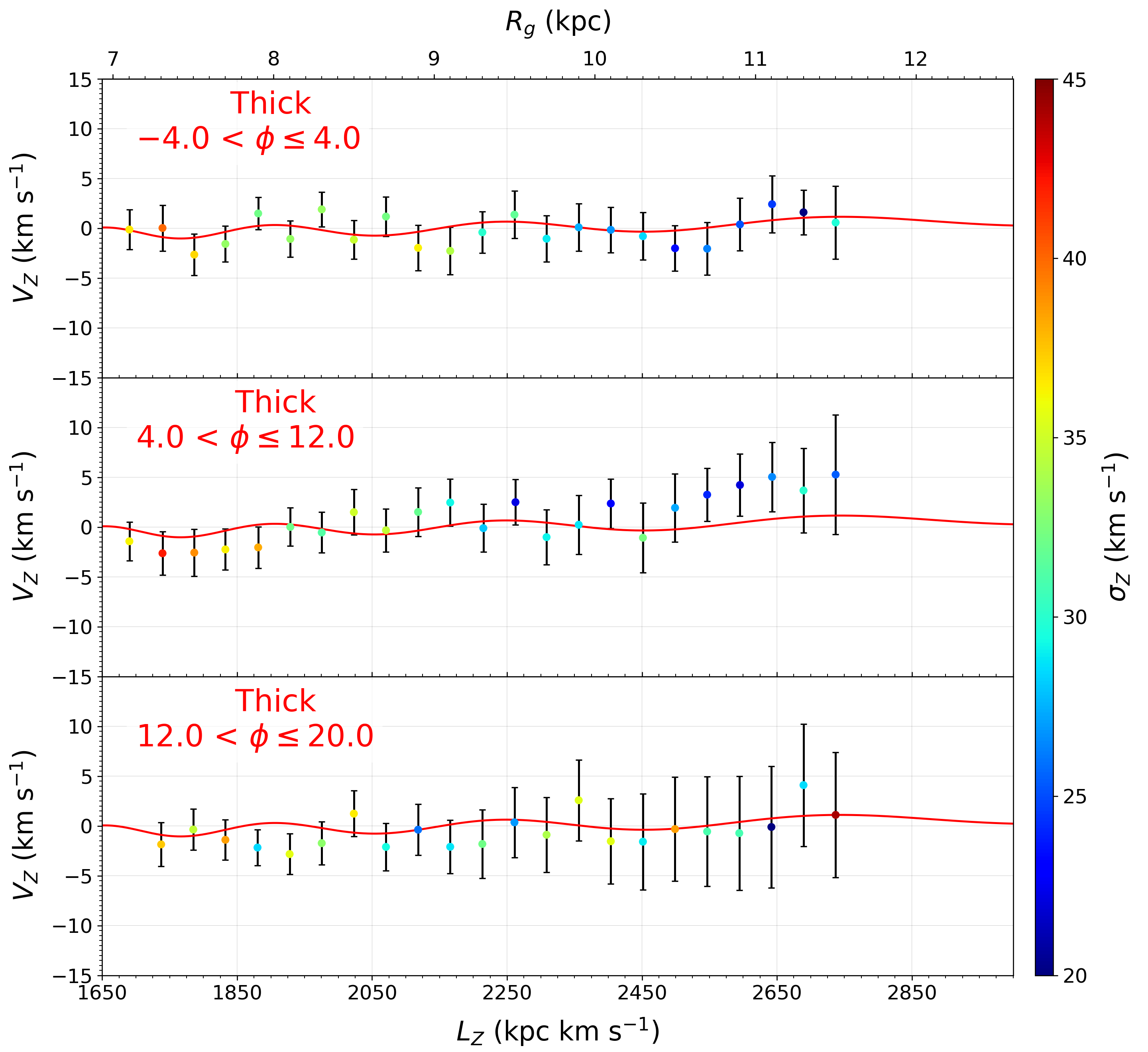}
}

\caption{Similar to Fig.\,2 but for the thick disk stars. The red lines are the best fits determined by Equation (2).}
\end{figure*}

The velocity dispersions ($\sigma$) in each bin are determined by 3$\sigma$ clipping to remove outliers. 
To ensure the accuracy of the stellar kinematic and orbital estimations, we further use conditions with stellar SNRs $>$ 20 and the distance uncertainty $\leq$ 10\%.
Those conditions ensure the measurement of the uncertainties of stellar 3D velocities are generally within 5.0\,km s$^{-1}$.
We also set $\ |V_{z}|$ $\leq$ 120 km s$^{-1}$ and [Fe/H] $\geq -1.0$ dex for the whole sample stars to exclude any possibility of contaminations of the halo stars \citep{Hayden2020, Sun2020}.
With the above selection, we finally selected 170,729 RC stars, of which 39,112 stars and 131,617 stars, respectively, from the APOGEE and LAMOST surveys.
The spatial distribution of the finally selected sample stars is shown in Fig.\,1.

To reveal more information about the disk structures and their evolutions from the $V_{Z}$--$L_{Z}$ relation, considering that the thin and thick disks are widely confirmed respectively as instead of the typical young and old populations \citep[e.g,.][]{Matteucci2001, Pagel2009, Haywood2013}, we further separate our sample stars to thin and thick disks by the distribution of these stars on the [Fe/H]--[$\alpha$/Fe] plane similar to Sun et al. ({\color{blue}{2023, 2024b}}), and finally, 135,009 stars and 23,168 stars are selected for thin and thick disks, respectively.
The properties of the two disks are shown in Table\,1.

\section{Results and Discussion}

Since the $V_{Z}$ from a warp will be a function of both $R$ and $\phi$ \citep[e.g.,][]{Poggio2017, Khanna2024, Sun2025}, and the onset radius of the warp is no less than 7\,kpc \citep[e.g.,][]{Chen2019, Uppal2024, Huang2024}, we study the $V_{Z}$--$L_{Z}$ relation at $L_{Z}$ $>$ 1650 kpc\,km\,s$^{-1}$ (corresponding to around $R_{g}$ $>$ 7\,kpc) in different $\phi$ bins for various populations.
The results are displayed in Fig.\,2--4, color-coded by their vertical velocity dispersions ($\sigma_{Z}$), with $R_{g}$ also labeled.

For the whole RC sample stars (Fig.\,3), the $V_{Z}$ as a function of $L_{Z}$ (or $R_{g}$) for different $\phi$ bins displays a similar shape, which shows an obvious increasing trend with a wave-like pattern as $L_{Z}$ (or $R_{g}$) increases.
In the Galactic-anti center direction ($-$4.0 $<$ $\phi$ $\leq$ 4.0 degree), the $V_{Z}$ increases steadily from $-$2.0\,km\,s$^{-1}$ at $L_{Z}$ $\sim$ 1700 kpc\,km\,s$^{-1}$ to 6.0\,km\,s$^{-1}$ at $L_{Z}$ $\sim$ 2850 kpc\,km\,s$^{-1}$.
The result at around 1750 $<$ $L_{Z}$ $<$ 2300 kpc\,km\,s$^{-1}$ (corresponding to around 7.35 $<$ $R_{g}$ $<$ 9.66\,kpc), is in good agreement with those from the TAGS sample \citep[][]{Schonrich2018} and the LAMOST-TGAS sample \citep[][]{Huang2018}, while there is a slight difference beyond this region.
Since the distance of their samples is determined by Gaia Parallax, their results have larger uncertainties compared to ours.
Furthermore, their results only cover a small portion of ours on the $L_{Z}$, which means we can find more complex structures of the Galactic disk.

For different $\phi$ bins, the increase in $V_{Z}$ with $L_{Z}$ is slightly higher in the $\phi$ bin with 4.0 $<$ $\phi$ $\leq$ 12.0 degree than that of other $\phi$ bins.
This implies that the line-of-node of the disk warp may be oriented at 4.0 $<$ $\phi$ $\leq$ 12.0 degree, which is in rough agreement with recent studies \citep[e.g.,][]{Huang2024, Sun2025}.

Previous studies provide a simple model to describe the relation of $V_{Z}$ as a function of $L_{Z}$ in the solar neighborhood \citep[e.g.,][]{Schonrich2018, Huang2018}, which is:
\begin{equation}
\label{eq:tiltangle2}
    V_Z = b + a L_Z + A\,\mathrm{sin}(2\pi c/L_Z + d)
\end{equation}
where $b$ + $a$\,$L_{Z}$ is used to describe the contribution from a well-known disk warp, and $A$\,sin(2$\pi c$/$L_{Z}$\,+\,$d$) is used to describe the wave-like pattern from disk waves.
Considering that $V_{Z}$ from a warp could be a function of both $R$ and $\phi$, we updated this equation as follows:
\begin{equation}
\label{eq:tiltangle2}
    V_Z = b + (a L_Z)\,\mathrm{cos}(\phi - \phi_0) + A\,\mathrm{sin}(2\pi c/L_Z + d)
\end{equation}
where $\phi_{0}$ is the LON of the disk warp.
Here we would like to clarify that several studies have confirmed that the LON is twisted and precessing with $R$ \citep[e.g.,][]{Dehnen2023, Cabrera-Gadea2024, Poggio2024}, but we still adopt a simple model that assumes a straight LON in this work, mainly because we aim to simultaneously detect both warp and waves signals from the $V_{Z}$--$L_{Z}$ relation.
The direction of the Sun–Galactic center represents $\phi$ = 0 degree.
However, we can find the $V_{Z}$--$L_{Z}$ at around $L_{Z}$ $>$ 2600 kpc\,km\,s$^{-1}$ (corresponding to round $R_{g}$ $\sim$ 11.0\,kpc) displays some deviations from the linear function shape.
This may be caused by the contribution from a well-known disk flare \citep[e.g.,][]{Hunter1969, Khoperskov2017, Sun2024b}, and the disk flare is significant in our result with the $\sigma_{Z}$ displaying obvious increases with $L_{Z}$ (see Fig.\,2).
Therefore, to provide a more accurate description of the $V_{Z}$--$L_{Z}$, we further update the linear function portion in Equation (2) to an exponential function, and then form an equation as follows:
\begin{equation}
\label{eq:tiltangle2}
    V_Z = b + (1 + B)^{L_Z}\,\mathrm{cos}(\phi - \phi_0) + A\,\mathrm{sin}(2\pi c/L_Z + d)
\end{equation}
where the $b$ + (1\,+\,$B$)$^{L_Z}$ represents the joint contribution of the disk warp and flare.

The best-fits are displayed by red lines in Fig.\,2, and the best-fit parameters are summarized in Table 2.
As the plot shows, the exponential function, $b$\,+\,(1\,+\,$B$)$^{L_{Z}}$, combined with a wave-like function, $A$\,sin(2$\pi$c/$L_{Z}$\,+\,$d$), can well describe the global trend of $V_{Z}$--$L_{Z}$ of the whole RC sample stars.
The amplitude of the wave-like pattern is around $A$ = 0.39 $\pm$ 0.15 kpc\,s$^{-1}$, which is slightly smaller than the results from the TGAS sample, with $A$ = 0.76 $\pm$ 0.07 kpc\,s$^{-1}$ \citep{Schonrich2018}, and the LAMOST-TGAS sample, with $A$ = 0.90 $\pm$ 0.14 kpc\,s$^{-1}$ \citep{Huang2018}.
It is crucial to highlight that our analysis deliberately excluded potential halo stars and eliminated outliers through a 3$\sigma$ clipping process for the $V_{Z}$ calculation.
Since the distance accuracy of our sample has improved several times compared to their results, this discrepancy in amplitudes may be attributed to the larger errors in their samples.
Moreover, their sample only provides reliable measurements in the solar neighborhood, this discrepancy may reflect the difference in the amplification of waves between the larger Galactic disk region and the local region.
Our result also yields the LON of the whole RC sample stars is $\phi_{0}$ = 5.68 $\pm$ 2.91 degree, this is in rough agreement with recent stellar chemistry result with $\phi_{0}$ = 9.22 $\pm$ 1.64 degree \citep[][]{Sun2025}.
It is worth noting that while multiple studies have shown that the LON is twisted and precessing with $R$ \citep[e.g.,][]{Chen2019, Dehnen2023, Cabrera-Gadea2024}, we still assume a straight LON in this work.
Therefore, when comparing our results with those reporting twisted and precessing LON,it is useful to compare our results to the range of LON values reported in those studies (instead of just a single number) in the approximate radial range covered by our dataset.
We find that our determined LON value falls within the range of values reported for a twisted and precessing LON in the approximate radial range covered by our data \citep[e.g.,][]{Dehnen2023, Poggio2024}, which indicates that our LON is in good agreement with other studies that show a twisted and precessing LON.

The thin disk stars exhibit a similar increasing trend in the $V_{Z}$--$L_{Z}$ relation as the whole sample stars in each $\phi$ bin.
For $-$4.0 $<$ $\phi$ $\leq$ 4.0 degree, the $V_{Z}$ increases from $-$2.0\,km\,s$^{-1}$ at $L_{Z}$ $\sim$ 1650 kpc\,km\,s$^{-1}$ to larger than around 6.0\,km\,s$^{-1}$ at $L_{Z}$ $\sim$ 2850 kpc\,km\,s$^{-1}$.
The increase in $V_{Z}$ with $L_{Z}$ is also slightly higher in the $\phi$ bin with 4.0 $<$ $\phi$ $\leq$ 12.0 degree than that of other $\phi$ bins.
The $V_{Z}$--$L_{Z}$ of different $\phi$ bins displays similar shapes, showing an exponential function with a wave-like pattern (see Fig.\,3), which means the features of disk waves, warp and flare appear in thin disk result, therefore, we also fit the thin disk result with Equation (3).
The best fit is plotted as the red dashed line in the upper panel of Fig.\,3, and the best-fit parameters are also summarized in Table 2.
The results yield the amplitude of the wave-like pattern of the thin disk to be around $A$ = 0.48 $\pm$ 0.12 kpc\,s$^{-1}$, and $\phi_{0}$  = 5.78 $\pm$ 2.89 degree.
Our determined thin disk LON is also in good agreement with recent stellar chemistry result with $\phi_{0}$  = 4.24 $\pm$ 1.64 degree \citep[][]{Sun2025} and young ($\sim$20--120 Myr) Cepheid sample result with $\phi_{0}$ = 6.14 $\pm$ 1.34 degree \citep[][]{Huang2024}.

The thick disk stars display a slightly increasing trend with a wave-like pattern in $V_{Z}$--$L_{Z}$.
For $-$4.0 $<$ $\phi$ $\leq$ 4.0 degree, the $V_{Z}$ increases from $-$0.5\,km\,s$^{-1}$ at $L_{Z}$ $\sim$ 1650 kpc\,km\,s$^{-1}$ to larger than 1.0$\sim$2.0\,km\,s$^{-1}$ at $L_{Z}$ $\sim$ 2650 kpc\,km\,s$^{-1}$.
Although the increasing trend in $V_{Z}$--$L_{Z}$ is extremely weak, we can also find the increase in $V_{Z}$ with $L_{Z}$ is also slightly higher in the $\phi$ bin with 4.0 $<$ $\phi$ $\leq$ 12.0 degree than that of other $\phi$ bins.
While the increasing trend in $V_{Z}$--$L_{Z}$ of each $\phi$ bin of the thick disk tends to show a linear-function shape.
Although this increasing trend is extremely weak, which also may point to the nature of the thick disk warp, and is in rough agreement with previous observations \citep[e.g.,][]{Li2020, Sun2024b}.
Considering that the thick disk has no flare behavior in our result (in Fig.\,4, the $V_{Z}$ of the thick disk stars displays no increasing trend with $L_{Z}$), and the wave-like pattern is also significant (see Fig.\,4), we use Equation (2) to fit the thick disk $V_{Z}$--$L_{Z}$ relations, and the best-fits are marked as the red dashed lines in the Fig.\,4, and the parameters that yield the best fit are detailed in Table 2.
The results yield the amplitude of the wave-like pattern of the thick disk is around $A$ = 0.60 $\pm$ 0.22 kpc\,s$^{-1}$, and $\phi_{0}$ = 4.10 $\pm$ 3.43 degree.

As discussed above, the wave-like pattern and disk warp appear in both young/thin and old/thick disks, we can confirm that the waves and warp are likely long-lived features in the Milky Way disk.
The long-lived feature of the disk warp means that it is potentially driven by secular processes like the disk perturbations from intergalactic magnetic fields \citep[e.g.,][]{Battaner1990} and a misaligned dark halo \citep[e.g.,][]{Sparke1988, Ostriker1989, Debattista1999}.

It is worth emphasizing that the thick disk result have larger uncertainties, the weak increasing trend in the $V_{Z}$--$L_{Z}$ of the thick disk may be influenced by contamination from thin disk stars.
We encourage further observational efforts with higher precision to address this issue.

\section{Conclusions}

In this study, we used a sample with over 170,000 RC stars selected from the LAMOST-APOGGE surveys to investigate the relation of $V_{Z}$ as a function of $L_{Z}$ (or $R_{g}$) across different $\phi$ bins for various populations. We find that:

1. The $V_{Z}$ of the whole RC sample stars shows a global increasing trend with a wave-like pattern as $L_{Z}$ (or $R_{g}$) increases for various $\phi$ bins, and is accurately described as $V_{Z}$ = $b$ + (1\,+\,$B$)$^{L_Z}$ cos($\phi$ $-$ $\phi_{0}$)+ $A$\,sin(2$\pi$c/$L_{Z}$ + $d$), where the $b$\,+ (1\,+\,$B$)$^{L_Z}$ reflects contributions from the disk warp and flare, and the $A$\,sin(2$\pi$c/$L_{Z}$\,+\,$d$) point to the presence of disk waves.
The amplitude of the wave-like pattern is $A$ = 0.39 $\pm$ 0.15 km\,s$^{-1}$, and the LON is oriented at $\phi_{0}$ = 5.68 $\pm$ 2.91 degree.

2. The $V_{Z}$ of thin disk stars displays a similar behavior to the whole sample stars, which is also well accurately described by the equation same to the whole sample stars, suggesting that the thin disk also exhibits features of disk warp, flare and waves.
The amplitude of the wave-like pattern is $A$ = 0.48 $\pm$ 0.12 km\,s$^{-1}$, and the LON is oriented at $\phi_{0}$ = 5.78 $\pm$ 2.89 degree.

3. The $V_{Z}$ of thick disk stars exhibits a slightly increasing trend with $L_{Z}$ (or $R_{g}$) for various $\phi$ bins, and shows a linear function with a wave-like shape, and is accurately well described as $V_{Z}$ = $b$ + $aL_{Z}$ cos($\phi$ $-$ $\phi_{0}$) + $A$\,sin(2$\pi$c/$L_{Z}$ + $d$), indicating the presence of disk warp and waves in the thick disk.
The amplitude of the wave-like pattern is $A$ = 0.60 $\pm$ 0.22 km\,s$^{-1}$, and the LON is oriented at $\phi_{0}$ = 4.10 $\pm$ 3.43 degree.

The disk warp and waves appear in both young/thin and old/thick disks indicating that they are likely long-lived features in the Milky Way disk, meaning that the disk warp is potentially driven by secular processes like the disk perturbations from intergalactic magnetic fields and a misaligned dark halo.

The LON of various populations displays a slight difference, with $\phi_{0}$ = 5.68 $\pm$ 2.91 degree for the whole RC sample stars, $\phi_{0}$ = 5.78 $\pm$ 2.89 degree for the thin disk stars, and $\phi_{0}$ = 4.10 $\pm$ 3.43 degree for the thick disk stars.

\section*{Acknowledgements}
We thank the anonymous referee for very useful suggestions to improve the work.
This work is supported by the NSFC projects 12133002, 11833006, and 11811530289, and the National Key R\&D Program of China No. 2019YFA0405500, 2019YFA0405503, and CMS-CSST-2021-A09, and the Postdoctoral Fellowship Program of CPSF under Grant Number GZC20240125, and the China Postdoctoral Science Foundation under Grant Number 2024M760240.

Guoshoujing Telescope (the Large Sky Area Multi-Object Fiber Spectroscopic Telescope LAMOST) is a National Major Scientific Project built by the Chinese Academy of Sciences. Funding for the project has been provided by the National Development and Reform Commission. LAMOST is operated and managed by the National Astronomical Observatories, Chinese Academy of Sciences.

\bibliographystyle{aasjournal}

\end{document}